\documentstyle[prl,twocolumn,aps,rotate,epsf,floats,amssymb,amsmath]{revtex}

\newcommand{\bq}{{\boldsymbol q}}
\newcommand{\bk}{{\boldsymbol k}}
\newcommand{\bl}{{\boldsymbol l}}

\newcommand{\bR}{{\boldsymbol R}}

\newcommand{\bz}{{\boldsymbol z}}
\newcommand{\by}{{\boldsymbol y}}
\newcommand{\bx}{{\boldsymbol x}}

\newcommand{\bOmega}{{\boldsymbol \Omega}}

\newcommand{\vF}{{v_{\rm F}}}

\def\rbx#1{\raisebox{-0.3ex}{$\scriptstyle #1$}}

\begin{document}
\setlength{\unitlength}{1cm}
\renewcommand{\arraystretch}{1.4}

\title{Low temperature physics of Sr$_2$RuO$_4$: multi-component 
superconductivity}
\author{Ralph Werner}

\address{Physics Department, Brookhaven National Laboratory, Upton, NY
11973-5000, USA}

\date{\today}

\maketitle

\centerline{Preprint. Typeset using REV\TeX}

\begin{abstract} 
It is shown that the properties of the superconductivity observed in
Sr$_2$RuO$_4$ can be accounted for by properly taking into
account the internal degrees of freedom of the system. The 
electrons are strongly correlated in the RuO$_2$ planes while the
superconducting phase transition is driven by ``umklapp scattering''
between the planes. The in-plane magnetic correlations are determined
via a non-perturbative approach. The orbital degeneracy and the
magnetic moments are internal degrees of freedom of the Cooper pairs
that account for the thermodynamic properties in the superconducting
phase. 
\end{abstract}
\pacs{PACS numbers: 74.20.-z, 74.70.-b, 75.50.-y}


With the discovery of the high temperature superconductors a whole
class of transition metal oxides became a focal point in condensed
matter research. These materials exhibit many unconventional
properties most of which are interpreted controversially to date. A
number of experimental probes\cite{IMK+98,LFK+98,DHM+00} show that the
paired electrons in Sr$_2$RuO$_4$ carry a magnetic moment making its
superconductivity unconventional\cite{MRS01}.  The present paper shows
that Sr$_2$RuO$_4$ is a striking example of how unconventional physics
can have conventional origins. 

Sr$_2$RuO$_4$ is the first layered transition metal oxide that
exhibits superconductivity in the absence of copper
ions\cite{MHY+94}. The lattice symmetry is tetragonal and
isostructural to La$_2$CuO$_4$ with lattice parameters $a=b=3.87$
{\AA} in the RuO$_2$ plane and $c=12.74$ {\AA} out-of-plane. In
contrast to La$_2$CuO$_4$ the ruthenate is a Fermi liquid in the
temperature range of $T_{\rm c} < T < 30$
K\cite{MYH+97,IMK+00}. Superconductivity appears below $T_{\rm c} \sim
1$ K.  

The significant correlations\cite{MHY+94,BJM+00} in Sr$_2$RuO$_4$, the
$S = 1$ moments on Ru$^{4+}$ impurities in Sr$_2$IrO$_4$, and
ferromagnetic correlations in SrRuO$_3$ led Rice and
Sigrist\cite{RS95} to propose that the superconducting order parameter
has $p$-wave symmetry promoted by ferromagnetic correlations analogous 
to $^3$He. A similar proposal was made by Baskaran\cite{Bask96} based
on a comparison with high $T_{\rm c}$ materials and emphasizing the
role of Hund's rule coupling. This idea is supported by
experiments that show that the static magnetic properties of
Sr$_2$RuO$_4$ are the same in the normal and the superconducting
phase\cite{IMK+98,DHM+00} while no conclusive experimental
proof\cite{LGL+00,TSN+01,ITY+01} for the  $p$-wave symmetry of the
superconducting order parameter was found. No indication for
ferromagnetic correlations has been found neither in neutron
scattering investigations\cite{SBB+99} nor other
approaches\cite{MPS00,DLS+00,SDL+01}. Furthermore, the specific
heat\cite{NMM00} and nuclear quadrupole resonance (NQR)\cite{IMK+00}
are consistent with two-dimensional gapless fluctuations in the
superconducting phase of Sr$_2$RuO$_4$ which are absent in superfluid
$^3$He. 

The present approach consists of three logical steps. First it is
shown that the correlations in the RuO$_4$ planes are dominated by
quantum fluctuations. Based on non-perturbative methods\cite{Wern02a}
it is then possible to account for the experimentally observed
magnetic structure factor\cite{SBB+99} in detail. 

In the second step it is shown that the coupling between the RuO$_4$
planes enhances the electron pair correlations via umklapp scattering 
processes\cite{Wern02b}. In contrast to the in-plane correlations the
inter-plane coupling can be treated perturbatively and leads to the
superconducting phase transition. The presence of the two degenerate
$d_{zx}$ and $d_{yz}$ orbitals together with Hund's rule coupling
leads to Cooper pairs that carry a magnetic moment.   

In a third step it is shown that the approach is consistent with a
multitude of experimental observations\cite{Wern02c}. The magnetic
moment of the Cooper pairs accounts for the static magnetic
properties\cite{IMK+98,DHM+00} of Sr$_2$RuO$_4$. Discussed closer
herein is the observed specific heat.


\paragraph*{Incommensurate magnetism}

In order to develop an understanding of the electronic correlations in
the RuO$_2$ planes it is instructive to first discuss how perturbative
approaches fail to describe the results from neutron
scattering\cite{SBB+99} in the normal phase.

The bare electronic band structure of Sr$_2$RuO$_4$ has been
determined unambiguously by a number of
methods\cite{BJM+00,MPS00,DLS+00}. Based on a tight binding band model
for the band structure the effects of the on-site Coulomb repulsion
can be included by a perturbative approach, the random phase
approximation (RPA). It is then possible to calculate the magnetic
structure factor. Various such studies have been performed with
similar results\cite{MS99,MTG01,EMJB01,Taki00}.

\begin{figure}[bt]
   \epsfxsize=0.48\textwidth
   \epsfclipon
   \centerline{\epsffile{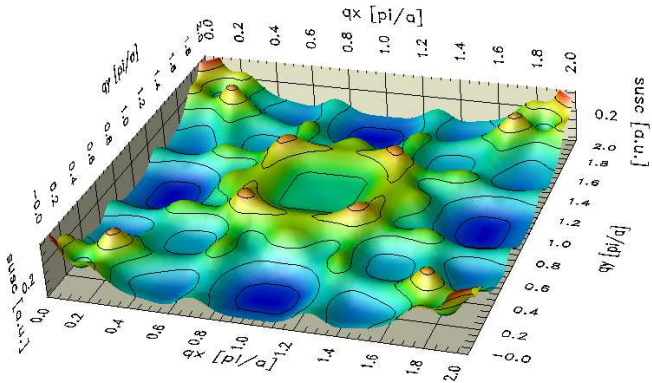}}
   \centerline{\parbox{\textwidth}{\caption{\label{Nfig1}
   \sl Typical plot of the magnetic susceptibility in the $q_x$-$q_y$
   plane from RPA. Data are folded with the experimental resolution in
   neutron scattering experiments.}}}      
   \end{figure}

Figure \ref{Nfig1} visualizes a typical result\cite{Wern02a} for the
imaginary part of the dynamical magnetic susceptibility in RPA based
on the band structures from the literature. The data have been
convoluted with the experimental resolution appropriate for inelastic
neutron scattering experiments. The structures in Fig.~\ref{Nfig1}
reflect the dispersion of the bands in the $q_x$-$q_y$ plane in
reciprocal space. While the results from neutron scattering show
pronounced peaks at $[\pm 0.6\frac{\pi}{a}, \pm 0.6\frac{\pi}{a}]$
many of the other prominent structures have not been 
observed\cite{SBB+99,Bradenprivate}.  

An obvious explanation is that the perturbative results underestimate
the quantum fluctuations of the quasi two-dimensional system. In the
presence of quantum fluctuations the magnetic susceptibility is
expected to be enhanced much more homogeneously. The question then
arises about the origin of the experimentally observed peaks at $[\pm
0.6\frac{\pi}{a}, \pm 0.6\frac{\pi}{a}]$.

The answer can be given in terms of the correlations of the $d_{zx}$
and $d_{yz}$ orbitals. The kinetic energy of the electrons in these
orbitals is dominantly one dimensional\cite{MS97,MPS00,MRS01}, namely
along $x$ for the $d_{zx}$ electrons and along $y$ for the $d_{yz}$
electrons. This allows us to bosonize the kinetic energy of the
electrons in these orbitals. The on-site Coulomb interaction can then
be included in a non-perturbative way. 

The Fermi velocity of the $d_{zx}$ and $d_{yz}$ electrons has been
estimated to be $\vF\approx 0.7$ eV$a$\cite{MPS00,SDL+01}. Perturbative 
approaches\cite{MS97,LL00,MTG01,EMJB01} involve values for the on-site
Coulomb repulsion that are of the same order as $\vF$, i.e., the 
system is rather in an intermediate than weak coupling regime. The
coupling hybridizes the $d_{zx}$ and $d_{yz}$ electrons leading to
dominantly quasi one-dimensional magnetic correlations along 
the diagonals of the system, i.e., along $x=\pm y$ or $q_x=\pm
q_y$. The mathematical derivation of those correlations from a
microscopic model is rather involved and the interested reader is
referred to a forthcoming article (Ref.\ \onlinecite{Wern02a}). 

\begin{figure}[bt]
   \epsfxsize=0.48\textwidth
   \epsfclipon
   \centerline{\epsffile{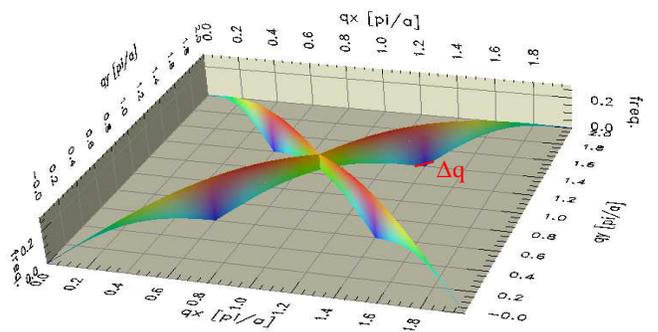}}
   \centerline{\parbox{\textwidth}{\caption{\label{Nfig2}
   \sl Effective, quasi one-dimensional magnetic excitation spectrum
   in the $q_x$-$q_y$ plane as anticipated from the $d_{zx}$ and
   $d_{yz}$ electronic subsystem at intermediate coupling.}}}      
   \end{figure}

The resulting anticipated quasi one-dimensional elementary magnetic
excitation spectrum in reciprocal space is depicted in Fig.\
\ref{Nfig2}.  The spectrum is gapless at $(\pm [\frac {2\pi}{a} - 2
k_{\rm F}], \pm[\frac {2\pi}{a} - 2 k_{\rm F}]) = (\pm
0.6\frac{\pi}{a}, \pm 0.6\frac{\pi}{a})$ which leads to a
dynamical magnetic structure factor that is strongly peaked as
observed experimentally\cite{SBB+99}. $k_{\rm F}$ is the Fermi wave
number of the $d_{zx}$ and $d_{yz}$ electrons. The existence of a
continuum is also consistent with experiments. Most striking though is
that the width of magnetic excitations at energy $\omega$ 
\begin{equation}\label{Deltaq}
   \Delta q = \frac{\omega}{v_{\rm eff}}
\end{equation}
is basically independent of the temperature as observed
experimentally\cite{SBB+99} (red bar Fig.\ \ref{Nfig2}). This feature
is difficult to explain within other theoretical approaches. The
comparison of Eq.\ (\ref{Deltaq}) with experiment yields a value of
the velocity of the elementary magnetic excitations of $v_{\rm eff}
\approx 120$ K which is consistent with the temperature dependence of
the peak intensities\cite{SBB+99,IMM+01}.  

In conclusion the intermediate coupling electronic correlations in the
RuO$_2$ planes of Sr$_2$RuO$_4$ must be described by a homogeneous,
two-dimensional part and quasi one-dimensional magnetic
correlations along the diagonals of the basal plane.


\paragraph*{Pair instability} 

For the system to undergo a phase transition the correlations in the
planes have to be coupled to form a three dimensional system. The
inter-plane hopping $t_\perp$ between the $d_{zx}$ and $d_{yz}$
orbitals has been estimated to be an order of magnitude smaller than the
in-plane hopping\cite{MS97,BJM+00}. We thus have a small
parameter justifying a perturbative approach. The inter-plane hopping
involving $d_{xy}$ orbitals is expected to be significantly smaller
for geometrical reasons\cite{BJM+00,MS97,MRS01} and is neglected.

Let us consider the lowest order coherent inter-plane pair hopping term   
\begin{equation}\label{Htperp2}
H_p = \sum_{\nu,\mu,\sigma \atop \nu',\mu',\sigma'}
          \frac{g^2_{\mu,\mu'} t_{\perp}^2}{\vF}
  \sum_{\bl, \bl'}\
         c^{\dagger}_{{\bl},\mu,\sigma} 
         c^{\phantom{\dagger}}_{{\bl}',\nu,\sigma}\,
         c^{\dagger}_{{\bl},\mu',\sigma'} 
         c^{\phantom{\dagger}}_{{\bl}',\nu',\sigma'}\,.
\end{equation}
The electron creation and annihilation operators are
$c^{\dagger}_{{\bl},\nu,\sigma}$ and
$c^{\phantom{\dagger}}_{{\bl},\nu,\sigma}$ for orbital $\nu \in
\{x,y\}$ with spin $\sigma \in \{\uparrow,\downarrow\}$ on site $\bl$.
The orbital indices $\nu,\mu,\nu',\mu' \in \{x,y\}$ are restricted by
the Pauli principle for $\sigma' = \sigma$ to $\mu' \neq \mu$ and
$\nu' \neq \nu$. 

The prefactor $1 \ge g^2_{\mu\neq\mu'} > g^2_{\mu =\mu'}$ takes into
account that the pair hopping into the same orbital is suppressed by the
Pauli exclusion principle. The presence of Hund's rule coupling favors
spin triplet pairs over spin singlet pairs. Consequently one expects
mixed orbital, spin-triplet pair hopping to be enhanced with respect
to mixed orbital, spin-singlet pair hopping and same orbital
hopping\cite{Wern02b,Taki00}. 

The Fermi operators in Eq.\ (\ref{Htperp2}) can be Fourier transformed
via   
\begin{equation}\label{FTFermi}
c_{{\bl},\nu,\sigma} = \frac{1}{\sqrt{N}}\sum_{\bk} 
 {\rm e}^{i\bk\bR_\bl}
         c_{{\bk},\nu,\sigma}\,. 
\end{equation}
The real-space positions of the Ru ions are given by $\bR_\bl$, the
$\bk$ are wave vectors. Limiting the discussion to the channel with
the largest pair hopping amplitude the Hamiltonian becomes in
reciprocal space 
\begin{equation}\label{HtperpFT}
H_p =  \sum_{b=x,y,z} \sum_{\bq} \frac{V_{\bq,t}}{N}\   
       P^\dagger_{t,b}(\bq) P^{\phantom{\dagger}}_{t,b}(\bq)\,.
\end{equation}
The mixed orbital, spin-triplet pair operators are defined by
\begin{equation}\label{tpair}
 P_{t,b}^{}(\bq)  = 
   \sum_{\sigma,\sigma',\sigma'',\bk}
         \sigma^y_{\sigma,\sigma''}\ \sigma^b_{\sigma'',\sigma'}\ 
   c^{}_{{\bq-\bk},x,\sigma}\,
                 c^{}_{{\bk},y,\sigma'} \,,
\end{equation}
where $\sigma^b_{\sigma,\sigma'}$ are the Pauli matrices with
$b\in\{x,y,z\}$. The effective potential is given by 
\begin{equation}\label{Veffa}
V_{\bq,t} = \frac{t_{\perp}^2}{\vF}\ 
     \cos \frac{a}{2}q_x \cos \frac{a}{2}q_y \cos \frac{c}{2}q_z\,.
\end{equation}
For $\bq = \bq_j$ with  
\begin{equation}\label{q_i}
\bq_j \in \left\{
\frac{2\pi}{a}\, \hat{\bx} + \frac{2\pi}{a}\, \hat{\by} + 
                                       \frac{2\pi}{c}\, \hat{\bz}, 
\frac{2\pi}{a}\, \hat{\bx}, \frac{2\pi}{a}\, \hat{\by}, 
                                \frac{2\pi}{c}\, \hat{\bz}
           \right\}
\end{equation}
the potential is extremal and attractive, i.e., $V_{\bq_j,t} \sim
-t_\perp^2/\vF$. As a consequence of the body centered
tetragonal lattice structure the $\bq_j$ are reciprocal lattice
vectors. In other words the body centered tetragonal lattice symmetry
allows for an umklapp scattering driven pair instability. 

The phase transition occurs at $T=T_{\rm c}$ when the Stoner criterion
\begin{equation}
\chi^{(P)}_{\rbx{t,b}}(\bq_j,0)\Big|_{T_{\rm c}} =
\chi^{(P)}_{\rbx{t,b}}(0,0)\Big|_{T_{\rm c}} = V_{0,2}^{-1}
\end{equation}
is fulfilled. $\chi^{(P)}_{\rbx{t,b}}(\bq,\omega)$ is the in-plane
pair-pair correlation function with the pair operators as given in
Eq.\ (\ref{tpair}). The strength of the effective potential can be 
estimated\cite{Wern02b} as $-V_{\bq_j,t} = V_{0,t} \sim t_\perp^2/\vF
\approx 6$ K which is consistent with the observed small critical
temperature $T_{\rm c} \sim 1.5$ K. In the absence of a magnetic
field and neglecting symmetry breaking effects $\chi^{(P)}_{\rbx{t,x}}
= \chi^{(P)}_{\rbx{t,y}} = \chi^{(P)}_{\rbx{t,z}}$ and below the phase 
transition all three order parameter components have a finite
expectation value $\langle P_{t,b}(0) \rangle \neq 0\ \forall\ b$.

The superconducting phase transition is induced by inter-plane umklapp
scattering. The Cooper pairs carry spin one that is the combined
magnetic moment of the two bound electrons. This is consistent with
the absence of a change of the Knight shift\cite{IMK+98} and of the
magnetic susceptibility\cite{DHM+00} in the superconducting phase.


\paragraph*{Order parameter fluctuations}

The order parameter in Sr$_2$RuO$_4$ has multiple internal degrees of
freedom. The magnetic degrees of freedom are given by fluctuations
between the three order parameter components $\langle P_{t,b=x,y,z} 
\rangle$. They can be parameterized by a SO(3) vector as $\bOmega_{\rm
s} = (\sin\theta_{\rm s}\sin\theta_{z}, \cos\theta_{\rm
s}\sin\theta_{z}, \cos\theta_{z})^\dagger$. 

Muon spin relaxation ($\mu$SR) measurements\cite{LFK+98} and the
out-of-plane critical field\cite{Wern02c} suggests that the magnetic
moment of the Cooper pairs lies predominantly in the $x$-$y$
plane\cite{Wern02b}. The vector $\bOmega_{\rm s}$ is then reduced to
SO(2) symmetry, i.e., $\theta_{z} = \pi/2$. This does not alter the
physical results discussed herein.

The mean-field approach as discussed in the previous section ignored 
possible spatial anisotropies of the order parameter. Critical field
measurements are consistent with the presence of two spatial order
parameter components\cite{Agte01,Wern02c}. The model discussed in the
context of the incommensurate magnetic fluctuations suggests that
there are two distinguished directions in the system. As becomes
apparent in Fig.\ \ref{Nfig2} those are along $q_x = q_y$ and $q_x =
-q_y$.  Indeed, it follows from the microscopic model that
the order parameter has two spatial components associated with these
directions\cite{Wern02b}. The components can be
parameterized by a two-component vector $\bOmega_{\rm f}$.  

In the superconducting (ordered) phase the fluctuations in each of the 
channels $\mu\in\{{\rm s},{\rm f}\}$ can be described by a
non-linear sigma model with the action\cite{Schu95}
\begin{equation}\label{Smu}
S_\mu = 
   \rho \int_{-L}^L\! d^2r \int_0^\beta d\tau 
       \!\sum_{\nu=x,y}\!  \Bigg[
\frac{(\dot{\bOmega}_{\mu})^2}{(v_{\nu})^{2}}
+ ( \partial_\nu \bOmega_{\mu})^2
\Bigg] .
\end{equation}
The correlations in Sr$_2$RuO$_4$ along the $z$ axis are two orders of
magnitude smaller than in the RuO$_2$
planes\cite{MYH+97,TSN+01}. The stiffness $\rho$ can 
thus be assumed to be isotropic in the plane but negligible along
$z$. The action Eq.\ (\ref{Smu}) is quasi two dimensional.

\begin{figure}[bt]
   \epsfxsize=0.44\textwidth
   \epsfclipon
   \centerline{\epsffile{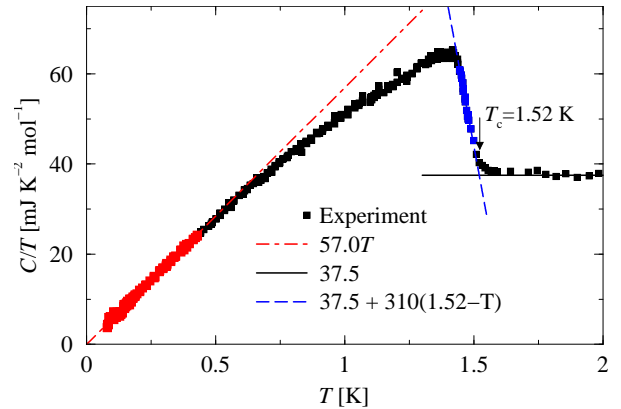}}
   \centerline{\parbox{\textwidth}{\caption{\label{Nfig3}
   \sl Specific heat. The dash-dotted line is a fit to the region of
   quadratic temperature dependence. The dashed line is the
   linear fit from Eq.\ (\ref{CsofT}).}}}
   \end{figure}

At low temperatures the action Eq.\ (\ref{Smu}) describes quasi
two-dimensional, gapless fluctuations. Their specific heat is
quadratic in temperature. The symbols in Fig.~\ref{Nfig3} show the 
experimental specific heat over temperature $C/T$ as published in
Ref. \onlinecite{NMM00} together with a linear low-temperature fit
(dash-dotted line).

For not too small magnetic fields the magnetic fluctuations described
by $S_{\rm s}$ in Eq.\ (\ref{Smu}) are suppressed and will not
contribute to the specific heat. One expects the slope of $C_H/T$
versus $T$ to be smaller than in the case without magnetic field
consistent with the experimental observation\cite{NMM00,Wern02c}.

A less obvious consequence of the gapless fluctuations in the
superconducting phase is the linear temperature dependence of the
specific heat below the phase transition. The blue symbols in Fig.\
\ref{Nfig3} show the data from Ref. \onlinecite{NMM00} in the linear
range near $T_{\rm c}$ together with a linear fit (dashed line). The
origin of this linearity can be understood by expressing the free
energy of the system in terms of the electronic excitation gap
$\Delta\sim |\langle P \rangle |$ in the superconducting
phase\cite{Tink96} valid for small reduced temperatures $t=1-T/T_{\rm
c} \ll 1$: 
\begin{equation}\label{FLandau}
F_{\Delta}  =  F_{0} - A\  t\ \Delta^2 + D\ \Delta^3  + B\ \Delta ^4 +
     {\rm O}(\Delta^5)\,.
\end{equation}
The presence of the third order term can be motivated by integrating
out the internal degrees of freedom $\bOmega_{\rm s}$ and
$\bOmega_{\rm f}$\cite{Wern02b}. Since Eq.\ (\ref{FLandau}) is an
expansion in $\Delta \ge 0$ and not in the order parameter itself the
third order term leads to a phase transition of third order in the
sense of Ehrenfest's definition.

The temperature dependence of the gap is obtained by minimizing the
free energy Eq.\ (\ref{FLandau}) as
\begin{equation}\label{DeltaofT}
\Delta(T)\big|_{t \ll 1} = \frac{2 A}{3 D}\ t\,.
\end{equation} 
The linear temperature dependence of the gap at the phase transition
has been observed via the $\mu$SR rate\cite{LFK+98,Wern02b} and the
excess current in tunneling experiments\cite{LGL+00,Wern02b}. In a
next step the specific heat near the phase transition
$C/T=-(\partial^2 F_\Delta)/(\partial \Delta^2)$ is found as 
\begin{equation}\label{CsofT}
\frac{C_s|_{t\ll 1}}{T_{\rm c}} = \frac{C_n}{T_{\rm c}} + 
\frac{8}{9}\frac{A^3}{T_{\rm c}^2 D^2}\ t
- {\rm O}(t^2)
\end{equation}
that is consistent with the linear temperature dependence at the phase
transition. $C_n$ is the specific heat in the normal phase. 

In conclusion it has been shown that the in-plane correlations of the 
RuO$_2$ subsystem of Sr$_2$RuO$_4$ exhibit dominant quantum
fluctuations making the application of perturbative approaches
questionable. The magnetic correlations have quasi one-dimensional
components along the diagonals of the basal plane of the crystal. The
superconducting instability is induced by inter-plane umklapp
scattering in the body centered crystal structure. The hopping 
amplitude is largest in the mixed orbital, spin-triplet
channel. Possible competing long range magnetic order is suppressed
because of the strong in-plane quantum fluctuations and the
frustrating body centered crystal structure.

The number of experimental results that are consistent with the
model for the superconductivity in  Sr$_2$RuO$_4$ presented here goes
beyond what can be discussed in this overview. The interested reader
is referred to Refs.\ \onlinecite{Wern02b} and
\onlinecite{Wern02c}. The new insight into the superconductivity in 
Sr$_2$RuO$_4$ may well help clarify the open questions posed by the
properties of other layered transition metal oxides.

I am grateful to V.\ J.\ Emery for initiating this
project. Instructive discussions with D.\ F.\ Agterberg, M.\ Braden,
S.\ T.\ Carr, J.\ M.\ Tranquada, A.\ M.\ Tsvelik, M.\ Weinert, and
B.\ O.\ Wells, are  acknowledged.


\end{document}